\documentclass[12pt]{article}%
\usepackage{amsmath}
\usepackage{amsfonts}
\usepackage{amssymb}
\usepackage{graphicx}%
\providecommand{\U}[1]{\protect\rule{.1in}{.1in}}
\begin{document}

\title{A Compact Approximate Solution to the Kondo Problem}
\author{Gerd Bergmann and Liye Zhang\\Department of Physics\\University of Southern California\\Los Angeles, California 90089-0484\\e-mail: bergmann@usc.edu}
\date{\today }
\maketitle

\begin{abstract}
A compact approximate groundstate of the Kondo problem is introduced. It
consists of four Slater states. The spin up and down states of the localized
d-impurity are paired with two localized s-electron states of opposite spin.
All the remaining s-electron states are rearranged forming two new optimal
orthonormal bases. Through a rotation in Hilbert space the two localized
states (and the rest of the bases) are optimized by minimizing the energy
expectation value. The ground-state energ$\;$y $E_{00}$ and the
singlet-triplet excitation energy $\Delta E_{st}$ are calculated numerically.
Although the two energies can differ by a factor of 1000, they are obtained
simultaneously. The singlet-triplet excitation energy $\Delta E_{st}$ is
proportional to $\exp\left[  -1/2J\rho\right]  $ and quite close to the Kondo
temperature $k_{B}T_{K}$. The cases for anti-ferromagnetic ($J>0$) and
ferromagnetic ($J<0$) coupling are investigated.


\end{abstract}

\section{Introduction}

The properties of magnetic impurities in a metal is one of the most
intensively studied problems in solid state physics. The work of Friedel
\cite{F28} and Anderson \cite{A31} laid the foundation to understand why some
transition-metal impurities form a local magnetic moment while others don't.
Kondo \cite{K8} showed that multiple scattering of conduction electrons by a
magnetic impurity yields a divergent contribution to the resistance in
perturbation theory. In the following three decades a large number of
sophisticated methods were applied to better understand and solve the Kondo
model, and it was shown that at zero temperature a Kondo impurity is in a
non-magnetic state. To name a few of these methods: scaling \cite{A51},
renormalization \cite{W18}, \cite{F30}, Fermi-liquid theory \cite{N14},
\cite{N5}, slave-bosons (see for example \cite{N7}), and the Bethe-ansatz
\cite{W12}, \cite{S29}. For a review see \cite{H20}. For numerical calculation
an approximate wave function by Varma and Yafet \cite{V7} which was later
extended to the large-spin limit \cite{G19}, \cite{B103} was particularly
productive. Finally after decades of research exact solutions of the Kondo and
Friedel-Anderson problems were derived \cite{W12}, \cite{A50}.

One of the authors \cite{B152}, \cite{B151} recently introduced a new approach
to the Friedel-Anderson impurity. This approach is based on building from the
s-electron spectrum a localized s-state for each spin which behaves as an
artificial Friedel resonance (AFR) states. This permits the construction of a
very compact approximate ground state (see appendix, equ. (\ref{hfa0})). It
requires solely the optimization of the two localized AFR states and consists
of a few Slater states (Slater determinants). This solution gives remarkably
good results for the ground-state energies and the occupation of the d-states.
The results are of the same quality as numerical calculation by Gunnarsson and
Schoenhammer \cite{G34} who applied the large $N_{f}$ method to the spin $1/2$
Friedel-Anderson model and included self-consistently up to $10^{7}$ basis states.

In this paper we introduce a similar ansatz to the Kondo effect. The intension
is to rederive some of the well known properties of the Kondo effect within a
rather simple and transparent frame work. The hope is to extend this rather
simple method to more complex problems such as the interaction between Kondo impurities.

\section{The AFR-Groundstate for the Kondo Effect}

The Kondo ground state which is suggested in this paper is derived from the
Friedel resonance problem. Below equation (\ref{hfa0}) gives the
Friedel-Anderson Hamiltonian. If one removes the Coulomb exchange interaction
term $Un_{d+}n_{d-}$ then one obtains the Hamiltonian for the Friedel
resonance. This is a single-particle Hamiltonian and the $n$-electron ground
state can be written as a simple product of single electron states which are a
hybridization of the d-electrons with the band s-electrons.

Due to the s-d-interaction between the s-electrons and the d-impurity the
s-electrons shift their weight towards the d-impurity. This is more clearly
expressed in a different version of the $n$-electron ground state of the
Friedel impurity \cite{B91}. The system forms from the s-band a state
$a_{0}^{\ast}$ which is localized in the close vicinity of the $d^{\ast}%
$-state. This $a_{0}^{\ast}$-state hybridizes with the $d^{\ast}$-state in the
form $\left(  Aa_{0}^{\ast}+Bd^{\ast}\right)  $, where $A,B$ are coefficients.
The remaining s-electrons are arranged in states $a_{i}^{\ast}$ ($1\leq i\leq
N-1$) which are orthogonal to $a_{0}^{\ast}$ and each other. (Furthermore the
matrix elements of the free electron Hamiltonian $H_{0}$ must be diagonal
$\left\langle a_{i}^{\ast}\Phi_{0}\left\vert H_{0}\right\vert a_{j}^{\ast}%
\Phi_{0}\right\rangle $ $=E\left(  i\right)  \delta_{ij}$ for \thinspace
$i,j\geq1,$ see appendix). Then the alternative version of the $\left(
n+1\right)  $-electron Friedel ground state is given by%

\begin{equation}
\psi_{Fr}=\left(  Aa_{0}^{\ast}+Bd^{\ast}\right)  \left\vert \mathbf{0}%
_{a}\right\rangle \label{psi_f}%
\end{equation}
where the hybrid $\left(  Aa_{0}^{\ast}+Bd^{\ast}\right)  $ plus the $n$
lowest states $a_{1}^{\ast}a_{2}^{\ast}..a_{n}^{\ast}=\left\vert
\mathbf{0}_{a}\right\rangle $ of the basis $\left\{  a_{i}^{\ast}\right\}  $
are occupied. The state $a_{0}^{\ast}$ can be found by numerical variation or
from an analytic expression (from $a_{0}^{\ast}$ the full basis $\left\{
a_{i}^{\ast}\right\}  $ can be derived). One of the authors has proved
\cite{B92} that (\ref{psi_f}) is an exact ground state of the Friedel impurity.

In the next step we consider the Friedel-Anderson impurity whose Hamiltonian
is given in equ. (\ref{hfa0}).
\begin{equation}
H_{FA}=\sum_{\sigma}\{\sum_{\nu=1}^{N}\varepsilon_{\nu}c_{\nu\sigma}^{\ast
}c_{\nu\sigma}+E_{d}d_{\sigma}^{\ast}d_{\sigma}+\sum_{\nu=1}^{N}V_{sd}%
(\nu)[d_{\sigma}^{\ast}c_{\nu\sigma}+c_{\nu\sigma}^{\ast}d_{\sigma}%
]\}+Un_{d+}n_{d-} \label{hfa0}%
\end{equation}
Its mean field solution is just a product of two Friedel solutions as given by
equ. (\ref{psi_f}), one for the spin up and the other for the spin down
electrons (with different effective $d^{\ast}$-state energies $E_{d\uparrow}$
and $E_{d\downarrow}$). It requires two localized s-states, which we denote
now as $a_{0}^{\ast}$ for spin up and $b_{0}^{\ast}$ for spin down and the
rest of their bases $\left\{  a_{i}^{\ast}\right\}  $ and $\left\{
b_{i}^{\ast}\right\}  .$ The two differ when the impurity is magnetic.

The mean field solution can be dramatically improved if one expands the
product of the two hybridized states $\left(  A_{\uparrow}a_{0\uparrow}^{\ast
}+B_{\uparrow}d_{\uparrow}^{\ast}\right)  \left(  A_{\downarrow}%
b_{0\downarrow}^{\ast}+B_{\downarrow}d_{\downarrow}^{\ast}\right)  $ and
assigns new coefficients $\left(  A,B,C,D\right)  $ for four two-electron
states. This yields%

\begin{equation}
\psi_{ms}=\left[  Aa_{0\uparrow}^{\ast}b_{0\downarrow}^{\ast}+Ba_{0\uparrow
}^{\ast}d_{\downarrow}^{\ast}+Cd_{\uparrow}^{\ast}b_{0\downarrow}^{\ast
}+Dd_{\uparrow}^{\ast}d_{\downarrow}^{\ast}\right]  \left\vert \mathbf{0}%
_{a\uparrow}\right\rangle \left\vert \mathbf{0}_{b\downarrow}\right\rangle
\label{psi_ms}%
\end{equation}
The optimal state $\psi_{ms}$ is obtained by varying the composition of
$a_{0}^{\ast}$ and $b_{0}^{\ast}$ and optimizing the coefficients $\left(
A,B,C,D\right)  $. This solution has the same structure as the mean field
solution but it yields a much lower ground-state energy. It also requires a
much larger threshold Coulomb exchange energy $U$ to form a magnetic moment at
the impurity.

The study of the Kondo effect taught us that the real ground state of the
Friedel-Anderson impurity is a symmetric state (often called a singlet state).
Such a state can be constructed from (\ref{psi_ms}) by reversing the spins of
all electrons and combining the two states. This yields
\begin{align}
\psi_{ss}  &  =\left[  Aa_{0\uparrow}^{\ast}b_{0\downarrow}^{\ast
}+Ba_{0\uparrow}^{\ast}d_{\downarrow}^{\ast}+Cd_{\uparrow}^{\ast
}b_{0\downarrow}^{\ast}+Dd_{\uparrow}^{\ast}d_{\downarrow}^{\ast}\right]
\left\vert \mathbf{0}_{a\uparrow}\right\rangle \left\vert \mathbf{0}%
_{b\downarrow}\right\rangle \label{psi_ss}\\
&  +\left[  A^{\prime}b_{0\uparrow}^{\ast}a_{0\downarrow}^{\ast}+C^{\prime
}b_{0\uparrow}^{\ast}d_{\downarrow}^{\ast}+Bd_{\uparrow}^{\ast\prime
}a_{0\downarrow}^{\ast}+D^{\prime}d_{\uparrow}^{\ast}d_{\downarrow}^{\ast
}\right]  \left\vert \mathbf{0}_{b\uparrow}\right\rangle \left\vert
\mathbf{0}_{a\downarrow}\right\rangle \nonumber
\end{align}
In equ. (\ref{psi_ss}) the creation operators have been reordered according to
their spin. By optimizing the localized states and the coefficients we
obtained an almost perfect agreement with Gunnarsson and Schoenhammer
\cite{G34} for the ground-state energy and the zero, single and double
occupation of the $d^{\ast}$-state.

In this paper we are interested in the Kondo effect. For the treatment of a
magnetic impurity Kondo used the exchange Hamiltonian $H_{sd}$ with the
exchange interaction $J_{\mathbf{k,k}^{\prime}}$. One generally approximates
the exchange interaction by a $\delta\left(  \mathbf{r}\right)  $-function:
$J\left(  \mathbf{r}\right)  =v_{a}\delta\left(  \mathbf{r}\right)  $ where
$v_{a}$ is atomic volume. Then the exchange (or Kondo) Hamiltonian has the form%

\begin{equation}
H_{sd}=v_{a}J\left[
\begin{array}
[c]{c}%
\left(  S_{+}\Psi_{\downarrow}^{\dag}\left(  0\right)  \Psi_{\uparrow}\left(
0\right)  +S_{-}\Psi_{\uparrow}^{\dag}\left(  0\right)  \Psi_{\downarrow
}\left(  0\right)  \right) \\
+S_{z}\left(  \Psi_{\uparrow}^{\dag}\left(  0\right)  \Psi_{\uparrow}\left(
0\right)  -\Psi_{\downarrow}^{\dag}\left(  0\right)  \Psi_{\downarrow}\left(
0\right)  \right)
\end{array}
\right]  \label{Hsd}%
\end{equation}
where $S_{+},S_{-},S_{z}$ are the spin operators of the impurity with spin
$S=1/2$ and $\Psi_{\uparrow}^{\dag}\left(  0\right)  $ and $\Psi_{\downarrow
}^{\dag}\left(  0\right)  $ represent field operators. The product $v_{a}%
J\Psi_{\sigma}^{+}\left(  0\right)  \Psi_{\sigma^{\prime}}\left(  0\right)  $
yields an energy since $\Psi_{\sigma}^{+}\left(  0\right)  \Psi_{\sigma
^{\prime}}\left(  0\right)  $ has the dimension of a density.

Schrieffer and Wolff \cite{S31} showed that there is an intimate connection
between the Friedel-Anderson and the Kondo impurity. The Friedel-Anderson
Hamiltonian can be transformed into an exchange Hamiltonian with an
anti-ferromagnetic exchange interaction $J>0$ (plus some additional terms). In
particular for large values of $U$ and large negative $E_{d}$ (for example
$E_{d}=-U$) they derived an effective $J$%
\begin{equation}
J\thickapprox-\left\vert V_{sd}\right\vert ^{2}\dfrac{U}{\left(
U+E_{d}\right)  E_{d}}>0 \label{Jef}%
\end{equation}

The Kondo impurity is in some respects a limiting case of the Friedel-Anderson
impurity. By increasing the exchange interaction $U$ (and decreasing
$E_{d}=-U/2$) the Friedel-Anderson impurity approaches the properties of a
Kondo impurity. For the Kondo impurity the d-impurity is a localized permanent
magnetic moment. This means that the $d^{\ast}$-state is always singly
occupied, either with spin up or down. We observe this transition in our
solution $\psi_{ss}$ (equ. \ref{psi_ss}). When we increase $U$ (with
$E_{d}=-U/2$) the coefficients $A,A^{\prime}$ and $D,D^{\prime}$ approach
zero. It becomes energetically to expensive to have an empty or doubly
occupied $d^{\ast}$-state. Therefore for the Kondo impurity we make the
following ansatz for the ground state%
\begin{align}
\psi_{K}  &  =\left(  Ba_{0\uparrow}^{\ast}d_{\downarrow}^{\ast}+Cd_{\uparrow
}^{\ast}b_{0\downarrow}^{\ast}\right)  \left\vert \mathbf{0}_{a\uparrow
}\mathbf{0}_{b\downarrow}\right\rangle +\left(  \overline{C}b_{0\uparrow
}^{\ast}d_{\downarrow}^{\ast}+\overline{B}d_{\uparrow}^{\ast}a_{0\downarrow
}^{\ast}\right)  \left\vert \mathbf{0}_{b\uparrow}\mathbf{0}_{a\downarrow
}\right\rangle \label{psi_K}\\
&  =B\psi_{B}+C\psi_{C}+\overline{C}\psi_{\overline{C}}+\overline{B}%
\psi_{\overline{B}}\nonumber
\end{align}
We remove the terms with zero and double d-occupancy from the AF ground state
$\psi_{ss}$.

We denote the state $\psi_{K}$ as the \textbf{AFR ground state} of the Kondo
effect since it uses the artificial Friedel resonance states $a_{0}^{\ast}$
and $b_{0}^{\ast}$.

As in the AF ground state $\psi_{ss}$ the states $a_{0}^{\ast}$ and
$b_{0}^{\ast}$ are localized s-states which are concentrated close to the
impurity. Their composition will be numerically optimized. Here, we sketch
briefly the construction of the $\left\{  a_{0}^{\ast},a_{i}^{\ast}\right\}  $
basis. (The construction of the $\left\{  b_{0}^{\ast},b_{i}^{\ast}\right\}  $
is completely analogous).

In this paper we use a finite s-electron band with $N$ states $c_{\nu}^{\ast}$
per spin and a magnetic $d^{\ast}$-state with spin $S=1/2$. For the present
calculations we use Wilson's band which extends from $-1$ to $+1$ and has an
exponentially fine energy spectrum close to the Fermi energy (see appendix).
The $a_{0}^{\ast}$-state is composed of these $c_{\nu}^{\ast}$-states%

\begin{equation}
a_{0}^{\ast}=\sum_{\nu=1}^{N}\alpha_{0}^{\nu}c_{\nu}^{\ast} \label{AFR}%
\end{equation}
(In the initial phase the coefficients $\alpha_{0}^{\nu}$ are arbitrarily
chosen, for example being constant).

The remaining s-electron basis$\left\{  c_{\nu}^{\ast}\right\}  $ has to be
made orthogonal to $a_{0}^{\ast}$. This yields the new remaining bases
$\left\{  a_{i}^{\ast}\right\}  $. The $a_{i}^{\ast}$ ($1\leq i\leq N-1$) are
built orthogonally to $a_{0}^{\ast}$ and to each other. In addition their
$(N-1)\times\left(  N-1\right)  $ sub-matrix of the s-band Hamiltonian
$H_{0}=\sum\varepsilon_{\nu}n_{\nu}$ is made diagonal. The $a_{i}^{\ast}$
together with $a_{0}^{\ast}$ represent a new basis. (Details of the
construction of the states $\left\{  a_{0}^{\ast},a_{i}^{\ast}\right\}  $ is
discussed in ref. \cite{B152}, \cite{B151} and briefly in the appendix). The
states $a_{i}^{\ast}$ are uniquely determined from the state $a_{0}^{\ast}$.
Their form is
\[
a_{i}^{\ast}=\sum_{\nu=1}^{N}\alpha_{i}^{\nu}c_{\nu}^{\ast}%
\]
The new bases $\left\{  a_{0}^{\ast},a_{i}^{\ast}\right\}  $ can be expressed
as a rotation of the original basis $\left\{  c_{\nu}^{\ast}\right\}  $ in
Hilbert space. The basis $\left\{  b_{0}^{\ast},b_{i}^{\ast}\right\}  $ is
built in complete analogy.

The free electron Hamiltonian can be expressed in either of the two new bases.
It is diagonal in the states $a_{i}^{\ast}$ for 1$\leq i\leq\left(
N-1\right)  $ and has the form%
\begin{equation}
H_{0}=\sum_{\nu=1}^{N}\varepsilon_{\nu}c_{\nu}^{\ast}c_{\nu}=\sum_{i=1}%
^{N-1}E\left(  i\right)  a_{i}^{\ast}a_{i}+E\left(  0\right)  a_{0}^{\ast
}a_{0}+\sum_{i=1}^{N-1}V_{fr}^{a}\left(  i\right)  \left[  a_{0}^{\ast}%
a_{i}+a_{i}^{\ast}a_{0}\right]  \label{afr0}%
\end{equation}

The Hamiltonian is equivalent to a Friedel Hamiltonian where $a_{0}^{\ast}$ is
an artificial Friedel resonance state (AFR state) which is purely composed of s-states.

After the initial construction of the two bases $\left\{  a_{0}^{\ast}%
,a_{i}^{\ast}\right\}  $ and $\left\{  b_{0}^{\ast},b_{i}^{\ast}\right\}  $
the AFR ground state (\ref{psi_K}) of the Kondo impurity can be formed.

\section{Numerical Results}

The procedure to obtain the optimal states $a_{0}^{\ast}$ and $b_{0}^{\ast}$
has been described in previous papers \cite{B91}, \cite{B152}, \cite{B151} and
is briefly sketched in the appendix. One starts from an s-band with $N$ states
possessing the energy $\varepsilon_{\nu}$. In all calculations the energy band
ranges from $-1$ to $1$ following Wilson's example. The density of states is
constant and equal to $\rho=1/2$ (corresponding to one state in the whole
band). This band is divided into $N$ adjacent cells $\left(  E_{\nu-1}:E_{\nu
}\right)  $. The energy values in the center (middle) of each cell yield the
spectrum $\varepsilon_{\nu}=\left(  E_{\nu}-E_{\nu-1}\right)  /2$.

It was pointed out by Wilson and will be confirmed in the present calculation,
that a very small energy spacing $\delta E$ at the Fermi energy is of
essential importance for obtaining the Kondo ground state. A linear energy
scale is therefore not well suited for the Kondo ground state because $\delta
E$ is always much too large. We use an energy spectrum with an exponential
energy scale. This energy spectrum was introduced by Wilson \cite{W18} in his
Kondo paper. Starting with the values of $-1$ and $+1$ the energy is $N/2$
times divided by 2, yielding the cells $\left(  -1:-1/2\right)  ,\left(
-1/2:-1/4\right)  ,$ .. $\left(  -1/2^{N/2}:0\right)  $. The negative value of
$\varepsilon_{\nu}$ are (for $1\leq\nu\leq N/2$): $-3/4,-3/8,..,-3/2^{\nu+1}%
,$.$.-3/2^{N/2+1},-1/2^{N/2+1}.$ The positive energies are the mirror image of
the negative ones.

From the chosen energy spectrum the two states $a_{0}^{\ast}$ and $b_{0}%
^{\ast}$ are constructed with initially arbitrary coefficients $\alpha
_{0}^{\nu}$ and $\beta_{0}^{\nu}$. The states $a_{0}^{\ast}$ and $b_{0}^{\ast
}$ determine uniquely the full bases $\left\{  a_{0}^{\ast},a_{i}^{\ast
}\right\}  $ and $\left\{  b_{0}^{\ast},b_{i}^{\ast}\right\}  $ (see
appendix). With the two bases the energy matrix elements between the different
Slater states are calculated.

The wave function has four components and the Hamiltonian consists of four
terms due to $H_{sd}$ and two terms due to $H_{0}.$ Therefore the energy
expectation value is composed of $96$ matrix elements. Many of these are
identical and one has to calculate only16 different matrix elements. They can
be transformed into determinants of $\left(  n+1\right)  \times\left(
n+1\right)  $ matrices. In the appendix some representetive matrix elements
are derived.

Since the two Slater states $\left(  \psi_{B},\psi_{\overline{C}}\right)  $
are not orthogonal (nor are $\left(  \psi_{C},\psi_{\overline{B}}\right)  $)
one has first to apply an transformation to the four-component basis $\left(
B,\overline{C},C,\overline{B}\right)  ^{\dag}$ before one can solve the
eigenvalue problem. (Details of the optimization are described in the appendix.)

The lowest value of $\Lambda$ yields the lowest energy expectation value
$\left\langle E\right\rangle $ for the chosen two bases $\left\{  a_{0}^{\ast
},a_{i}^{\ast}\right\}  $ and $\left\{  b_{0}^{\ast},b_{i}^{\ast}\right\}  $.
Then states $a_{0}^{\ast}$ and $b_{0}^{\ast}$ are rotated in Hilbert space
until the value $\left\langle E\right\rangle $ reaches a minimum. The
resulting state is defined as the AFR ground state of the Kondo impurity and
the energy $E_{00}=\left\langle E\right\rangle $ is its ground-state energy.%
\begin{align*}
&
{\includegraphics[
height=3.1756in,
width=3.9767in
]%
{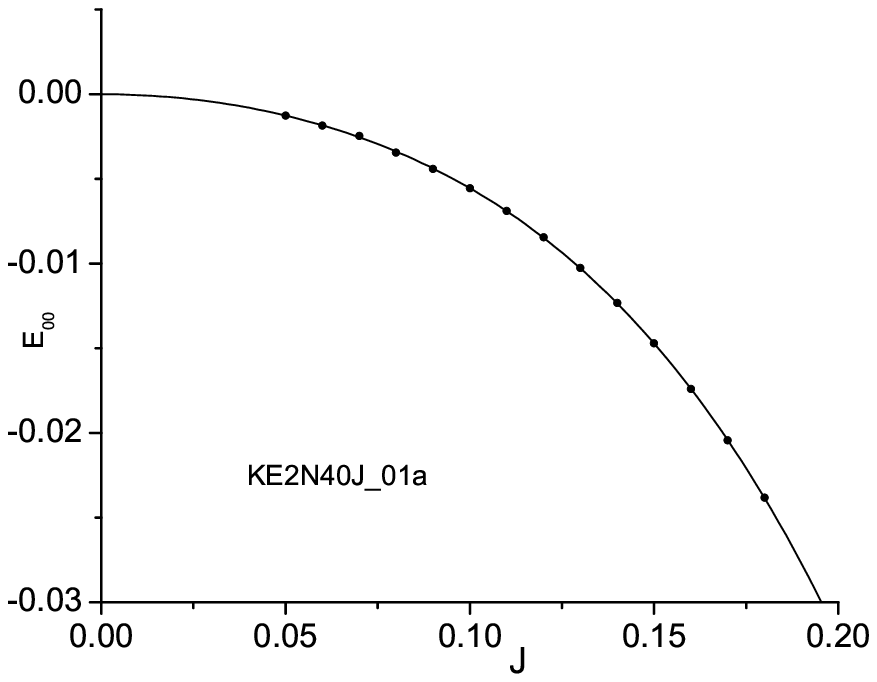}%
}%
\\
&
\begin{tabular}
[c]{l}%
Fig.1: The energy of the AFR ground state of the Kondo\\
impurity as a function of the exchange interaction $J$ for\\
$N=40$ states.
\end{tabular}
\end{align*}

This ground-state energy $E_{00}$ is plotted in Fig.1 as a function of the
exchange interaction $J$. The curve shows the results for the exponential
spectra with $N=40$. The corresponding value for the energy spacing at the
Fermi energy is $\delta E=1/2^{20}$ $\thickapprox1\times10^{-6}$.

The ground-state energy $E_{00}$ can be well expressed by the following dependence%

\[
E_{00}\left(  J\right)  =-.5017\ast J^{2}+.435\ast J^{3}-9.65\ast J^{4}4
\]

This ground-state energy is \textbf{not }the Kondo energy. It is sometimes
called the perturbational part of the ground-state energy. It is (in most
cases) ignored in the Kondo effect as the non-singular part of the
ground-state energy. Below we will compare this ground state with the
ground-state energy for ferromagnetic coupling, i.e. for negative values of
$J$.

The solution of the eigenvalue equation (\ref{ewp}) yields four eigenvalues.
The lowest eigenvalue corresponds to the ground-state energy. This should be a
singlet state. We expect that the next eigenvalue of the energy, corresponding
to the first excited state, represents a triplet state. To confirm these
expectance we calculate the expectation value of the total spin squared
$\mathbf{S}^{2}=\left(
{\textstyle\sum_{i}}
\mathbf{s}_{i}\right)  ^{2}$ for the two states. For $J=0.1$ we find in the
ground state $\left\langle \mathbf{S}^{2}\right\rangle =0.04$. For the first
excited state we obtain for the square of the total spin $\left\langle
\mathbf{S}^{2}\right\rangle =1.99$. This is very close to the value 2 which we
expect for $S=1$. We conclude that the ground state is not a perfect singlet
state but is pretty close to it. It is essentially a state with $S=0,$ but it
has a small component of a larger total spin such as $S=1,2..$. Similarly the
first excited state has essentially the total spin $S=1$. But it too has a
small admixture of states with $S$ equal to $0,2,..$. (The fact that
$\left\langle \mathbf{S}^{2}\right\rangle $ is so close to $2$ means that the
contributions of $S=0$ and $S=2$ essentially cancel each other). For the rest
of the paper we denote the ground state as the singlet state $\psi_{K,0}$. The
first excited state we denote as a triplet state $\psi_{K,1}$.

The energy difference between these two lowest energy states we denote as the
singlet-triplet excitation energy $\Delta E_{st}$. This energy difference
$\Delta E_{st}$ is very small and $\ln\left(  \Delta E_{st}\right)  $ is shown
in Fig.2 in a logarithmic plot as a function of the inverse coupling strength
$1/J$. $\Delta E_{st}$ follows an exponential law covering six orders of
magnitude in the range $0.05\leq J\leq0.18.$ The linear full curve in Fig.2 is
given by $\Delta E\thickapprox5\ast D\ast\exp\left[  -x\right]  $ where
$x=1/\left(  2J\rho_{0}\right)  $. The other data and curves in Fig.2 are
explained in the discussion.%

\begin{align*}
&
{\includegraphics[
height=3.2179in,
width=3.9219in
]%
{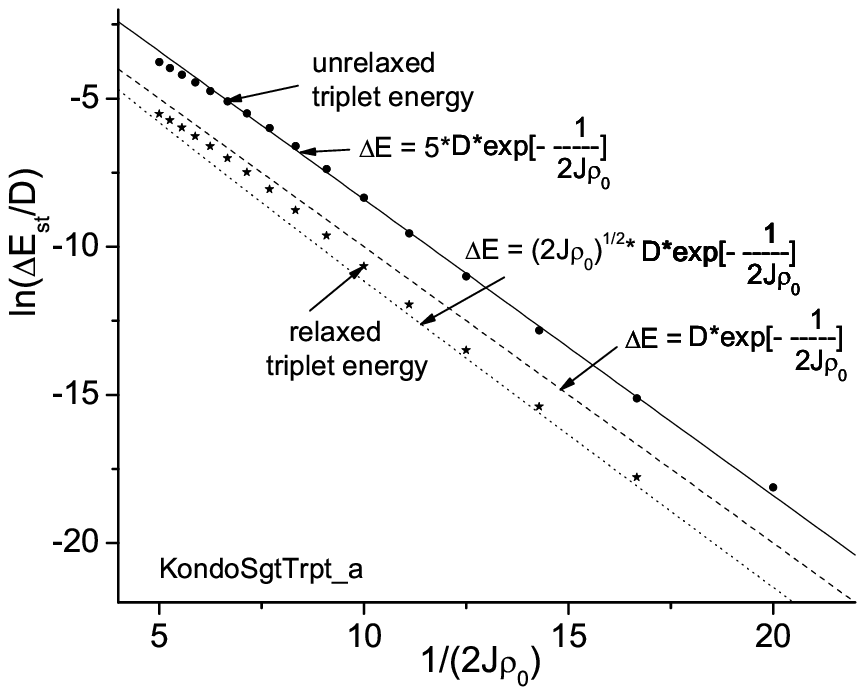}%
}%
\\
&
\begin{tabular}
[c]{l}%
Fig.2: The energy difference between the singlet and triplet\\
state. The full circles are obtained as the energy difference\\
between the ground state and the first excitated state. The\\
stars and the dashed and dotted curves are explained in the\\
discussion. A exponential Wilson band with $N=40$\\
and $N=60$ is used in the calculation.
\end{tabular}
\end{align*}%
\[
\]

Wilson has shown that the energy separation of the states at the Fermi energy
has to be smaller than the Kondo energy to obtain the full Kondo effect (or in
Wilson's words: to make the transition into the infinitely strong coupling
case). This same behavior is observed in our AFR solution.

We investigate how critical the smallest level separation $\delta E$ at the
Fermi energy is. For this purpose we vary the number $N$ of states for the
exponential energy spectrum. The level spacing at the Fermi energy $\delta E$
depends exponentially on $N$: $\delta E$=2$^{-N/2}.$For each $N$ we calculate
the AFR ground state for two values of $J,$ $J=0.1$ and $J=0.07$. In Fig.3 the
singlet-triplet energy difference $\Delta E_{st}$ is plotted versus the
smallest energy spacing $\delta E$ in a log-log plot. The number $N$ of energy
levels is varied from $N=12$ (at the right side) to $N=48$ (left side). One
recognizes that for sufficiently small $\delta E$ (on the left side of the
figure) the singlet-triplet excitation energy assumes a constant value. With
increasing $\delta E$ the excitation energy increases. The dashed straight
line has a slope of $0.85$ corresponding to an increase of $\Delta E_{st}$
proportional to $\left(  \delta E\right)  ^{0.85}$. Fig.3 demonstrates that
the transition to a constant $\Delta E_{st}$ occurs roughly for $\delta
E\thickapprox\Delta E_{st}$ and that $\delta E$ must be at least a factor 10
smaller than the final result for $\Delta E_{st}$ to give a reliable value for
$\Delta E_{st}$.
\begin{align*}
&
{\includegraphics[
height=3.3167in,
width=3.9618in
]%
{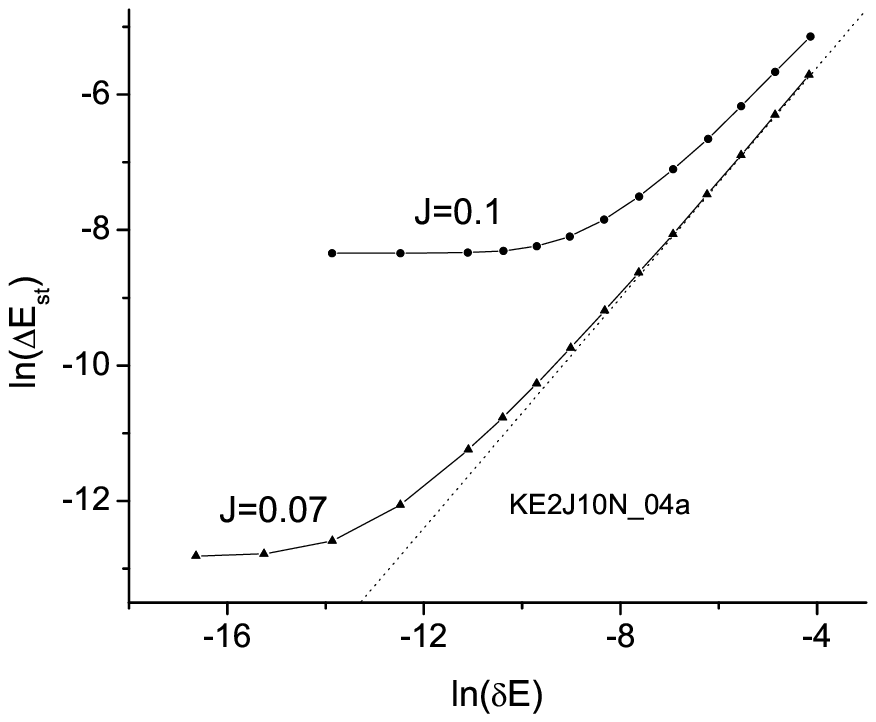}%
}%
\\
&
\begin{tabular}
[c]{l}%
Fig.3: The logarithm of the singlet-triplet excitation energy versus\\
the logarithm of smallest energy spacing $\delta E$ at the Fermi energy.
\end{tabular}
\end{align*}

\[
\]

\subsection{AFR ground state for ferromagnetic coupling}

One of the interesting aspects of the exchange Hamiltonian is that it only
yields the Kondo anomaly for anti-ferromagnetic coupling or $J>0$. We apply
the AFR-ansatz to the case of ferromagnetic coupling, i.e. $J<0$. In Fig.4 the
resulting ground-state energy is plotted versus the absolute value $\left\vert
J\right\vert $. The ground-state energy lies somewhat above that of the Kondo
case with anti-ferromagnetic coupling. The difference is about $10\%$ for
small $\left\vert J\right\vert $ and a factor two$\ $for $\left\vert
J\right\vert \thickapprox0.18)$.

It is in particular interesting that the two lowest eigenvalues of equation
(\ref{ewp}) differ on by about $10^{-11}$ for sufficiently small level
separation $\delta E$. Here the value of $\left\vert \Delta E_{st}\right\vert
$ is always much smaller than $\delta E$. In other words, at all temperatures
the two states, singlet and triplet, are degenerate and the impurity is
magnetic. The AFR solutions for anti-ferromagnetic and ferromagnetic coupling
are distinctively different.
\begin{align*}
&
{\includegraphics[
height=3.2179in,
width=4.0473in
]%
{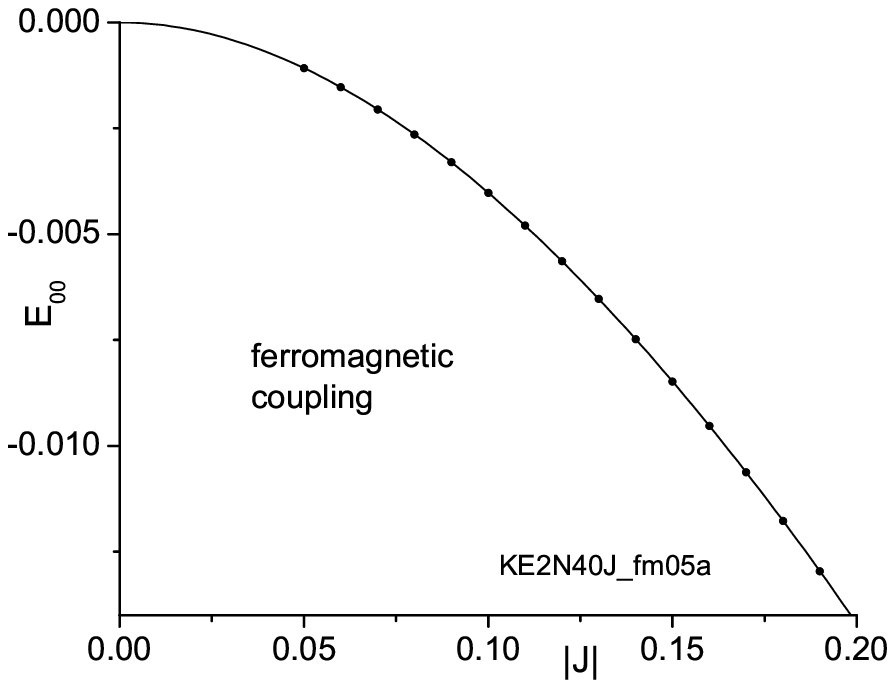}%
}%
\\
&
\begin{tabular}
[c]{l}%
Fig.4: The ground-state energy of the system for negative $J,$\\
i.e., for ferromagnetic coupling.
\end{tabular}
\end{align*}

\[
E_{0mf}=-.4630\ast J^{2}+.680\ast J^{3}-.697\ast J^{4}%
\]%
\[
\]

\subsection{The artificial Friedel resonance states}

\subsubsection{Anti-ferromagnetic coupling $\left(  J>0\right)  $}

The whole bases $\left\{  a_{0}^{\ast},a_{i}^{\ast}\right\}  $ and $\left\{
b_{0}^{\ast},b_{i}^{\ast}\right\}  $ can be derived from the composition of
the AFR states $a_{0}^{\ast}$ and $b_{0}^{\ast}$. In Fig.5a,b the coefficients
$\alpha_{0}^{\nu}$ and $\beta_{0}^{\nu}$ of the states $a_{0}^{\ast}$ and
$b_{0}^{\ast}$ are plotted versus the state index $\nu$ for $N=40$ (using
equ.(\ref{AFR})). On the left and right side of $\nu=N/2$ the values of $\nu$
and $\left(  N-\nu\right)  $ represent essentially the logarithm of the
energy. The coefficients are always positive because this yields the maximal
interaction of the AFR states with the magnetic $d^{\ast}$ state. The state
$a_{0}^{\ast}$ is mainly composed of states below $\varepsilon_{F}$ and
$b_{0}^{\ast}$ has its main weight at energies above $\varepsilon_{F}$.
However, in the center close to the Fermi energy the amplitudes of the two
localized states are essentially identical. This is a behavior which we also
observed for the Friedel-Anderson impurity.

\begin{align*}
&
\begin{tabular}
[c]{ll}%
{\includegraphics[
height=2.2208in,
width=2.7829in
]%
{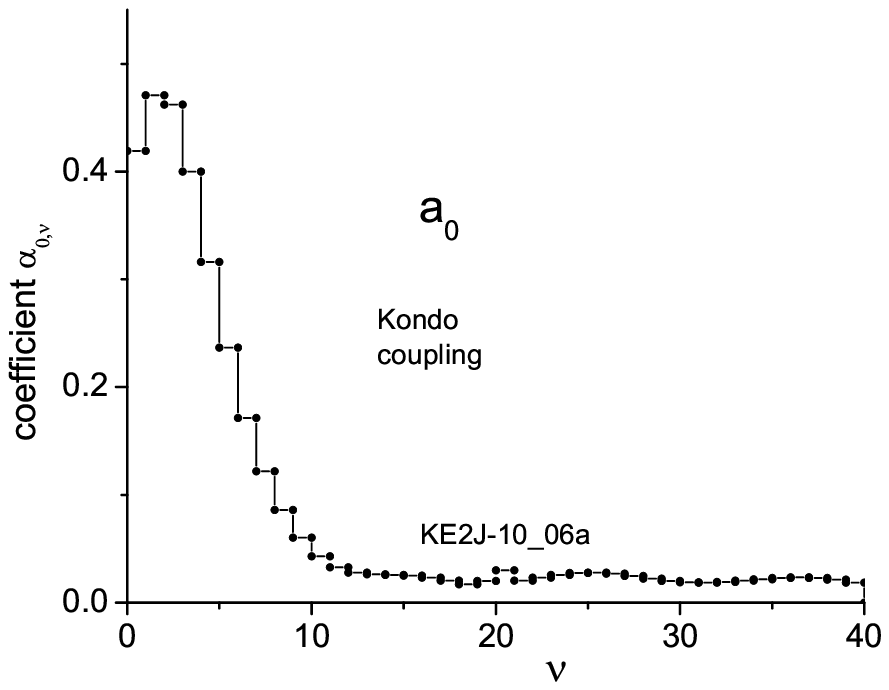}%
}%
&
{\includegraphics[
height=2.2208in,
width=2.7829in
]%
{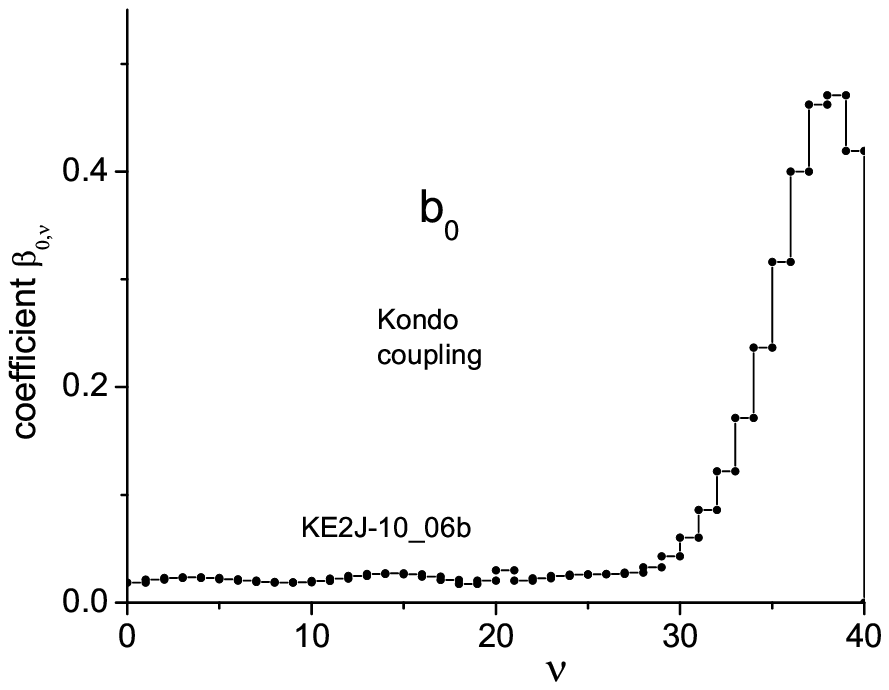}%
}%
\end{tabular}
\ \ \ \ \ \\
&
\begin{tabular}
[c]{l}%
Fig.5a,b: The coefficients of the AFR states $\alpha_{0}^{\nu}$ and $\beta
_{0}^{\nu}$ as a function of $\nu$ for the\\
two AFR states $a_{0}^{\ast}$ and $b_{0}^{\ast}$. The exchange interaction is
for anti-ferromagnetically\\
coupling ($J=0.1).$%
\end{tabular}
\end{align*}

\[
\]

For large energies the amplitudes $\alpha_{0}^{\nu}$ and $\beta_{0}^{\nu}$ of
$a_{0}^{\ast}$ and $b_{0}^{\ast}$ are rather different. For the analysis at
small energies we plot the occupation density $\left\vert \alpha_{0}^{\nu
}\right\vert ^{2}/\left(  E_{\nu}-E_{\nu-1}\right)  $ and $\left\vert
\beta_{0}^{\nu}\right\vert ^{2}/\left(  E_{\nu}-E_{\nu-1}\right)  $ as a
function of $\nu$. This is shown in Fig.6 where both occupations are plotted
in the same figure. At energies close to the Fermi energy the occupations of
$a_{0}^{\ast}$ and $b_{0}^{\ast}$ are almost identical.
\[%
{\includegraphics[
height=3.1199in,
width=3.9892in
]%
{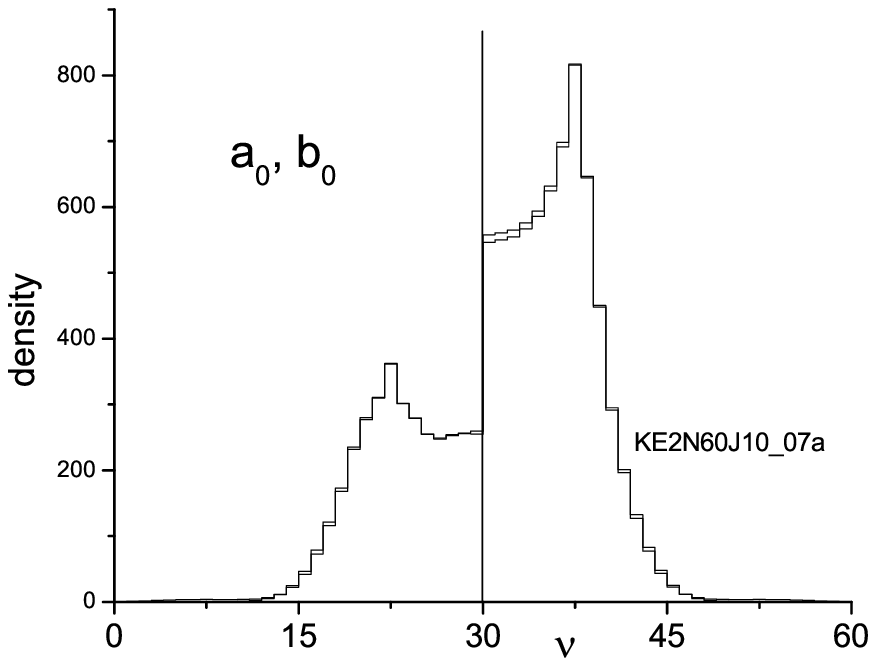}%
}%
\]
Fig.6: The occupation of the $a_{0}^{\ast}$ and $b_{0}^{\ast}$ AFR states for
$N=60$. Note that the figure shows two curves as a function of $\nu$ $\left(
0\leq\nu\leq60\right)  $. At small energies (close to the center arrow) the
two are almost identical.

\subsubsection{Ferromagnetic coupling $\left(  J<0\right)  $}

For comparison we plot the coefficients $\alpha_{0}^{\nu}$ and $\beta_{0}%
^{\nu}$ for $J=-0.1$, i.e. the case of ferromagnetic coupling in Fig.7a,b. For
the ferromagnetic coupling we do not observe the two maxima as in the Kondo
case. Furthermore, the coefficients have negative sign in half of the energy
range, reducing the amplitudes of the AFR states at $r=0$, the position of the
$d^{\ast}$-moment$.$%
\begin{align*}
&
{\includegraphics[
height=2.3404in,
width=2.9323in
]%
{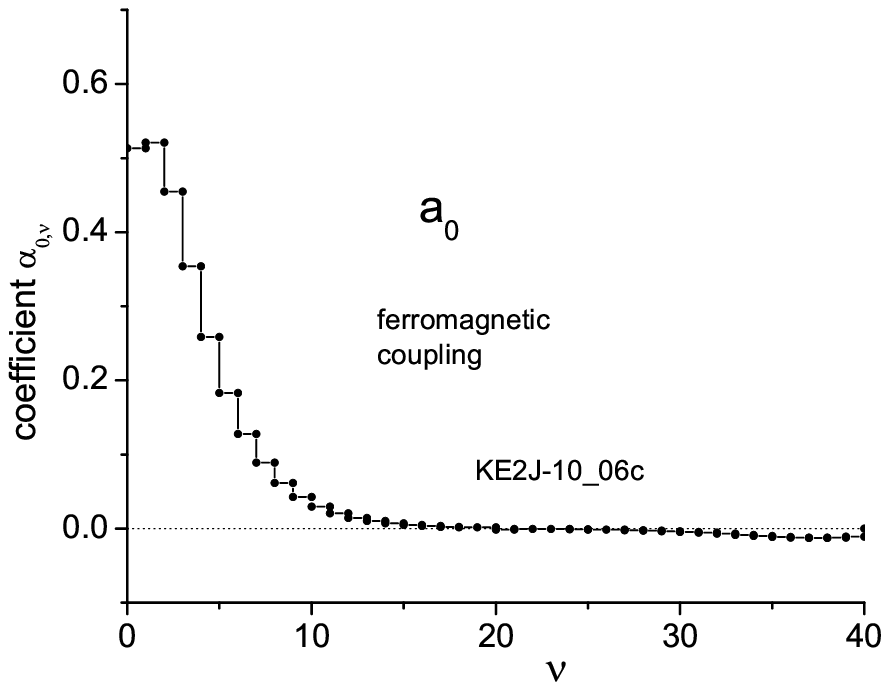}%
}%
{\includegraphics[
height=2.347in,
width=2.9406in
]%
{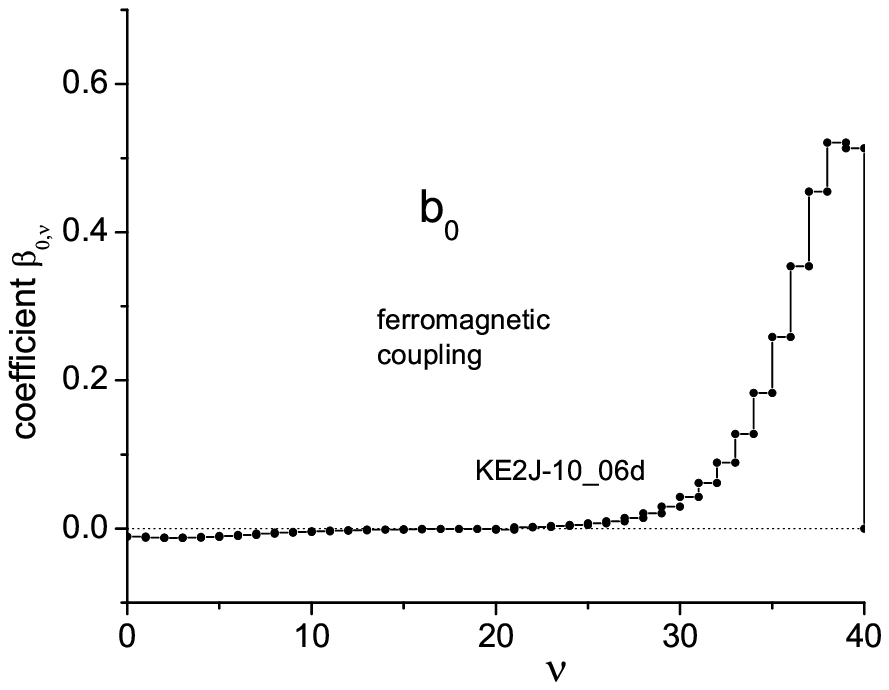}%
}%
\\
&
\begin{tabular}
[c]{l}%
Fig.7a,b: The coefficients of the AFR states $\alpha_{0}^{\nu}$ and $\beta
_{0}^{\nu}$ as a function of $\nu$ for the two\\
AFR states $a_{0}^{\ast}$ and $b_{0}^{\ast}$. The exchange interaction is
ferromagnetically coupling ($J=-0.1).$%
\end{tabular}
\end{align*}%
\[
\]

\section{Discussion}

Yosida \cite{Y2} introduced in the 1960's an approximate solution for the
Kondo Hamiltonian of the form%
\[
\Psi=\dfrac{1}{\sqrt{2}}\left(  a_{0\uparrow}^{\ast}d_{\downarrow}^{\ast
}+d_{\uparrow}^{\ast}a_{0\downarrow}^{\ast}\right)  \left\vert 0\right\rangle
\]
where
\[
\left\vert 0\right\rangle =%
{\textstyle\prod\limits_{\substack{\sigma,\mathbf{k}\\k<k_{F}}}}
c_{\mathbf{k,\sigma}}^{\ast}\Phi_{0}%
\]
is the quasi-vacuum with all states $c_{\mathbf{k}}^{\ast}$ below the Fermi
energy occupied and $a_{0}^{\ast}$ is an optimized single electron state which
is composed of states with $k>k_{F}$. The state $a_{0}^{\ast}$ is therefore
orthogonal to all occupied states. $\Phi_{0}$ is the vacuum state. Yoshida's
ground-state energy and singlet-triplet energy were of the same order of
magnitude with
\[
E_{00}=D\exp\left(  -\frac{1}{\frac{3}{2}J\rho}\right)
\]

Yosida's ansatz and our suggested AFR ground state are related to each other
in may respects: (i) Both use a localized state $a_{0}^{\ast}$, (ii) both form
pair states between the localized state and the $d^{\ast}$ state with zero
spin component in z-direction, (iii) both fill the remaining electrons into
states which are orthogonal to $a_{0}^{\ast}$. The main difference is that in
our approach the state $a_{0}^{\ast}$ together with a second state
$b_{0}^{\ast}$ are composed of all band states. This requires the remaining
electron states to be aligned perpendicular to the localized states. Our AFR
ground state is in a way a revival of Yosida's approach.

\subsubsection{Ferromagnetic versus anti-ferromagnetic coupling}

The numerical calculations show that the ground state for the
anti-ferromagnetic ($J>0$, Kondo effect) and the ferromagnetic coupling
($J<0$) are very different.

\begin{itemize}
\item The ground-state energy for positive $J$ lies lower. The behavior is
roughly described by the two fits to the numerical results:

\begin{itemize}
\item For $J>0$: $E_{0+}=-.5017\ast J^{2}+.435\ast J^{3}-9.65\ast J^{4}$

\item For $J<0$: $E_{0-}=-.4630\ast\left\vert J\right\vert ^{2}+.680\ast
\left\vert J\right\vert ^{3}-.697\ast\left\vert J\right\vert ^{4}$
\end{itemize}

\item There is essentially no singlet-triplet excitation energy in the case of
ferromagnetic coupling. The reason is that the system adjusts in such a way
that not only the states with opposite $d^{\ast}$-state spin (for example
$\psi_{B}$ and $\psi_{C}$) are orthogonal but even the states with the same
$d^{\ast}$-state spin orientation (for example $\psi_{B}$ and $\psi
_{\overline{C}}$) are quasi-orthogonal. This becomes visible in the
(multi-electron) scalar products between the state $\psi_{B},\psi
_{\overline{C}},\psi_{C},\psi_{\overline{B}}$. For $J=-0.1$ the scalar
products $\left\langle \psi_{B}|\psi_{\overline{C}}\right\rangle $ $\ $and
$\left\langle \psi_{C}|\psi_{\overline{B}}\right\rangle $ are of the order of
$10^{-6}$ while in the anti-ferromagnetic Kondo ($J=0.1)$ $\left\langle
\psi_{B}|\psi_{\overline{C}}\right\rangle $ $\ $and $\left\langle \psi
_{C}|\psi_{\overline{B}}\right\rangle $ are of the order of $1/3$. In the case
of the ferromagnetic coupling $J<0$ the two bases $\left\{  a_{0}^{\ast}%
,a_{i}^{\ast}\right\}  $ and $\left\{  b_{0}^{\ast},b_{i}^{\ast}\right\}  $
align themselves in such a way that the scalar product between $\left\vert
0_{a}\right\rangle $ and $\left\vert 0_{a}\right\rangle $ with $n$ occupied
states is of the order of $\left\langle 0_{a}|0_{b}\right\rangle
\thickapprox10^{-3}$. The bases align themselves quasi-orthogonal. Therefore
the reduction of the ground-state energy is solely due to z-component of the
interaction. The energy matrix element of the x-y-component of the interaction
is practically zero.
\end{itemize}

\subsubsection{Coefficients of the ground- and excited state}

Our Kondo-ground state $\psi_{K}$ is given by (\ref{psi_K}). The
coefficient-vector $\left(  V\right)  =\left(
\begin{array}
[c]{cccc}%
B & \overline{C} & C & \overline{B}%
\end{array}
\right)  ^{\dag}$ contains interesting information about the structure of the
states. Here we discuss the results for $J=0.1$, $N=40$. In the ground state
$\left(  V\right)  ^{\dag}$ takes the values $\left(  V\right)
_{\text{ground}}^{\dag}=$ (.703,.0626,.0626,.703) and in the excited state we
find $\left(  V\right)  _{\text{excited}}^{\dag}=$ (.705,-0635,0635,-.705).
(Note that state $\psi_{K,0}$ and $\psi_{K,1}$ are normalized but not the
vectors $\left(  V\right)  $ because the pairs of states $\psi_{B}%
,\psi_{\overline{C}}$ and $\psi_{C},\psi_{\overline{B}}$ are not orthogonal as
discussed in the appendix). For both, ground and excited states, the
sub-states with $a_{0}^{\ast}$ have a large amplitude while the sub-states
with the state $b_{0}^{\ast}$ have a small amplitude. In the ground state all
coefficients are positive. In the excited state the sub-states with opposite
$d^{\ast}$-spin have opposite sign. These are the signatures of a singlet
ground state and triplet excited state. (Note that the states are ordered
according to their spin).

For the ferromagnetic coupling we obtain for $J=-0.1$, $N=40$ the following
coefficients: In the ground state $\left(  V\right)  ^{\dag}$ takes the values
$\left(  V\right)  _{\text{ground}}^{\dag}=$ (-.706,.-.0400,.0400,.706). For
the essentially degenrated excited state we find $\left(  V\right)
_{\text{excited}}^{\dag}=$ (.706,.-.0400,-.0400,.706). This means that in the
ground state the components with $Ba_{0\uparrow}^{\ast}d_{\downarrow}^{\ast}$
and $\overline{B}d_{\uparrow}^{\ast}a_{0\downarrow}^{\ast}$ form a triplet
state (as do the components $Cd_{\uparrow}^{\ast}b_{0\downarrow}^{\ast}$ and
$\overline{C}b_{0\uparrow}^{\ast}d_{\downarrow}^{\ast}$) but the signs of the
states which contain $a_{0}^{\ast}$ and $b_{0}^{\ast}$ with the same spin are opposite.

\subsubsection{Singlet-triplet excitation energy}

The triplet state $\psi_{K,1}$ is obtained by minimizing the ground-state
energy. Then $\psi_{K,1}$ is the first excited state. This state $\psi_{K,1}$
is composed of the same Slater states $\psi_{B},\psi_{C},\psi_{\overline{B}%
},\psi_{\overline{C}}$ as the singlet state. Only the coefficients, and
particularly their signs, are different than in $\psi_{K,0}$. But this state
is not necessarily the triplet state with the lowest energy. To obtain the
lowest triplet state during the minimization we impose on the coefficients
$B,C$ and $\overline{B},\overline{C}$ the conditions
\[%
\begin{array}
[c]{ccc}%
\overline{B}=-B &  & \overline{C}=-C
\end{array}
\]
This automatically enforces the triplet state. With these conditions we repeat
the optimization procedure to obtain the lowest triplet energy. The resulting
state is a relaxed triplet state. We denote it as $\overline{\psi_{K,1}}$. Its
Slater states $\psi_{B},\psi_{C},\psi_{\overline{B}},\psi_{\overline{C}}$ are
different than for the singlet ground state $\psi_{K,0}$. Its energy
$\overline{E_{K,1}}$ is lower than the energy $E_{K,1}$ of the unrelaxed
triplet state $\psi_{K,1}$.

Now we can take the energy difference between the energy of the relaxed
triplet state and the singlet state. Because the two values are obtained in
two different optimization calculations the absolute accuracy of the
optimization has to be better than $10^{-10}$ (in units of the band width). We
plot the new excitation energy in Fig.2 as stars. It lies below the unrelaxed
excitation energy. The latter is approximately given by $\Delta E_{st}%
\thickapprox5D\exp\left[  -1/\left(  2J\rho_{0}\right)  \right]  $. Our new
values for the excitation energy lie between two theoretical curves: (i)
$\Delta E_{st}=D\exp\left[  -1/\left(  2J\rho_{0}\right)  \right]  ,$given by
the dashed curve and (ii) $\Delta E_{st}=\sqrt{2J\rho_{0}}D\exp\left[
-1/\left(  2J\rho_{0}\right)  \right]  $, given by the dotted curve. Both
expressions are given in the literature as approximate values for the Kondo
temperature $k_{B}T_{K}$. The numerical values lie closer to the second expression.

\subsection{The artificial Friedel resonance states}

In the Fig.5a,b and 7a,b the coefficients $\alpha_{0}^{\nu}$ (or $\beta
_{0}^{\nu}$ respectively) as they occur in equ. (\ref{AFR}) are plotted as a
function of $\nu$. One has to recall that for $N=40$ the Fermi energy lies in
the center at $\nu=20$. The energies above the Fermi energy are: $\frac{1}%
{2}2^{-19},\frac{3}{2}2^{-19},\frac{3}{2}2^{-18},..\frac{3}{2}2^{-1}.$ The
energy states below the Fermi energy are a mirror image of the state above.

The Kondo and the ferromagnetic case have in common that the coefficients of
$a_{0}^{\ast}$ and $b_{0}^{\ast}$ are almost mirror images. The state
$a_{0}^{\ast}$ has its main weight at negative energies and $b_{0}^{\ast}$ at
positive energies.

Beyond that the ferromagnetic and anti-ferromagnetic cases differ
considerably. The coefficients in the Kondo case are all positive whereas in
the ferromagnetic case the signs in the positive and negative energy regimes
are different.

\section{Conclusion}

In this paper we have developed a compact (approximate) ground state for the
Kondo problem. Two localized s-electron states, $a_{0}^{\ast}$ and
$b_{0}^{\ast}$ are built from the s-band. Their amplitudes, normally two times
40 numbers, determine fully the ground state. This ground state consists of
four Slater determinants. It yields two energies, the total ground-state
energy $E_{00}$ and the singlet-triplet excitation energy $\Delta E_{st}$.
While $E_{00}$ depends essentially quadratically on the coupling constant $J,$
the energy $\Delta E_{st}$ shows an exponential dependence. It is given by%
\[
\Delta E_{st}\thickapprox5\ast\exp\left(  -\frac{1}{2J\rho}\right)
\]
and is proportional to the Kondo temperature.

The energy scales of $\Delta E_{st}$ and $E_{00}$ are very far apart. They
differ by a factor 100 to 1000. It is remarkable that our solution works for
both energy ranges at the same time.

Our ansatz shows also a very different solution for the Kondo ($J>0$) case
than for the ferromagnetically coupled case ($J<0$). In particular the latter
has no singlet-triplet splitting, and the different sub-states $\psi_{X}$
avoid each other by forming quasi-orthogonal multi-electron states.

The structures of the localized AFR states $\alpha_{0}^{\ast}$ and
$b_{0}^{\ast}$ differ considerably between the $J>0$ and the $J<0$ case. In
particular for the $J>0$ case the coefficients have two maxima in an
intermediate positive and negative energy range. A better understanding of
these structures is left for future investigations.

Acknowledgement: The research was supported by the National Science Foundation
DMR-0439810.\newpage

\appendix

\section{Appendix}

\subsection{Wilson's s-electron basis}

Wilson \cite{W18} in his Kondo paper considered an s-band with energy values
ranging from $-1$ to 1. In the next step Wilson replaced the continuum of
s-states by a discrete set of states. This is done on a logarithmic scale. The
discrete energy values are 1, $1/\Lambda$, $1/\Lambda^{2}$, etc and $-1$,
$-1/\Lambda$, $-1/\Lambda^{2}$, etc where $\Lambda$ is a parameter larger than
one. (In this paper $\Lambda=2$ is chosen). These discrete $\xi_{\nu}$ points
are used to define a sequence of intervals: the interval $\nu$ (for $\nu$%
$<$%
N/2) is $\xi_{\nu-1}=-1/2^{\nu-1}<$ $\varepsilon$ $<-1/2^{\nu}=\xi_{\nu}$
(there are equivalent intervals for positive $\xi$-values where $\nu$ is
replaced by $\left(  N-\nu\right)  $ but we discuss here only the negative
energies). The new Wilson states $c_{\nu}^{\ast}$ are a superposition of all
states in the energy interval $\left(  \xi_{\nu-1},\xi_{\nu}\right)  $ and
have an (averaged) energy $\left(  \xi_{\nu}+\xi_{\nu-1}\right)
/2=\allowbreak\left(  -\dfrac{3}{2}\right)  \dfrac{1}{2^{\nu}}$, i.e.
$-\frac{3}{4},-\frac{3}{8},-\frac{3}{16},..,-\frac{3}{2\cdot2^{N/2}},-\frac
{1}{2\cdot2^{N/2}}.$ This spectrum continues symmetrically for positive
energies. The essential advantage of the Wilson basis is that it has an
arbitrarily fine energy spacing at the Fermi energy.

While Wilson chose the logarithmic energy scale for his scaling approach, in
this paper the physical background is somewhat different. We start (like
Wilson) with the somewhat artificial energy band in the range $\left(
-1,1\right)  $ which has a constant density of states. The volume of the host
is $V_{h}$ and the atomic volume $v_{A}.$ The full band (per spin) shall have
one electron per atomic volume. So the total number of electrons of a full
band in the volume $V_{h}$ is $Z=V_{h}/v_{A}$. The electron density per spin
of a full band is $n=Z/V_{h}=1/v_{A}$. The density of states (per spin) is
$Z/2$ in the volume $V_{h}$ and equal to $\rho=1/2$ in the volume $v_{A}$.

In this model the original energy levels have a (constant) separation of
$\delta E=2v_{A}/V_{h}$ which will be in the range between $10^{-8}$ to
$10^{-23}$. In an energy interval $\left(  \xi_{\nu-1},\xi_{\nu}\right)  $ one
has a large number $Z_{\nu}$ of electron states $\varphi_{\mu}\left(
r\right)  $ with $1\leq\mu\leq Z_{\nu}$. Their level separation shall be
constant and the value of the wave function at the origin $\varphi_{\mu
}\left(  0\right)  =1/\sqrt{V_{h}}$ is independent of $\mu$.

Next we form a new basis $\chi_{\alpha}\left(  r\right)  $ for the energy
range $\left(  \xi_{\nu},\xi_{\nu-1}\right)  $. We renumber all the original
states $\varphi_{\mu}\left(  \mathbf{r}\right)  $ in this energy range from
$1$ to $Z_{\nu}$ and define the new state (or wave function) $\chi_{\alpha
}\left(  \mathbf{r}\right)  $ $\left(  1\leq\alpha\leq Z_{\nu}\right)  $%
\[
\chi_{\alpha}\left(  \mathbf{r}\right)  =\frac{1}{\sqrt{Z_{\nu}}}%
{\textstyle\sum_{\mu=1}^{Z_{\nu}}}
\varphi_{\mu}\left(  \mathbf{r}\right)  e^{i2\pi\alpha\mu/Z_{\nu}}%
\]
In the new basis $\chi_{\alpha}\left(  \mathbf{r}\right)  $ all states except
$\chi_{Z_{\nu}}\left(  r\right)  $ have a zero amplitude at the origin.
\[
\chi_{\alpha}\left(  0\right)  =\left\{
\begin{array}
[c]{ccc}%
=0 & \text{for} & 1\leq\alpha<Z_{\nu}\\
\sqrt{\frac{Z_{\nu}}{V_{h}}} & \text{for} & \alpha=Z_{\nu}%
\end{array}
\right\vert
\]

In the following we denote the state $\chi_{Z_{\nu}}$ as $\psi_{\nu}$ and
neglect all other $\chi_{\alpha}\left(  \mathbf{r}\right)  $-states because
the latter do not interact with the impurity. The field operator $\Psi\left(
\mathbf{r}\right)  $ in the original basis $\varphi_{\mu}$ has the form
($c_{\varphi_{\mu}}$ is the annihilation operator for the state $\varphi_{\nu
}\left(  \mathbf{r}\right)  $)
\[
\Psi\left(  \mathbf{r}\right)  =%
{\textstyle\sum_{\mu=1}^{Z}}
\varphi_{\mu}\left(  \mathbf{r}\right)  c_{\varphi_{\mu}}%
\]
We can divide $\Psi\left(  \mathbf{r}\right)  $ into one part which is
non-zero at $\mathbf{r=0}$ and a second part which vanishes at zero. For the
interaction with the d-impurity only the first part contributes. This part is
equal to
\[
\overline{\Psi}\left(  \mathbf{r}\right)  =%
{\textstyle\sum_{\nu}}
\psi_{\nu}\left(  \mathbf{r}\right)  c_{\nu}%
\]
At $\mathbf{r=0}$ one has
\begin{align*}
\Psi\left(  \mathbf{0}\right)   &  =\overline{\Psi}\left(  \mathbf{0}\right)
=%
{\textstyle\sum}
\psi_{\nu}\left(  \mathbf{0}\right)  c_{\nu}=%
{\textstyle\sum}
\sqrt{\frac{Z_{\nu}}{V_{h}}}c_{\nu}\\
&  =\frac{1}{\sqrt{v_{a}}}%
{\textstyle\sum}
\sqrt{\frac{Z_{\nu}}{Z}}c_{\nu}=\frac{1}{\sqrt{v_{a}}}%
{\textstyle\sum}
\sqrt{\frac{\Delta\zeta_{\nu}}{2}}c_{\nu}%
\end{align*}

The width of the energy ranges can be chosen arbitrarily. We use Wilson's
exponential energy spectrum. This resulting basis $\left\{  c_{\nu}^{\ast
}\right\}  $ is the starting point for all calculations.

Since $J$ is the matrix element per atomic volume we have to use also the
density of states per atomic volume. The latter is
\[
\rho=\frac{1}{2}%
\]

\subsection{Construction of the Basis $a_{0}^{\ast}$, $a_{i}^{\ast}$}

For the construction of the state $a_{0}^{\ast}$ and the rest of basis
$a_{i}^{\ast}$ one starts with the s-band electrons $\left\{  c_{\nu}^{\ast
}\right\}  $ which consist of $N$ states (for example Wilson's states). The
$d^{\ast}$-state is ignored for the moment. \newline

\begin{itemize}
\item In step (1) one forms a normalized state $a_{0}^{\ast}$ out of the
s-states with:
\end{itemize}

\begin{equation}
a_{0}^{\ast}=\sum_{\nu=1}^{N}\alpha_{\nu}^{0}c_{\nu}^{\ast}%
\end{equation}
The coefficients $\alpha_{\nu}^{0}$ can be arbitrary at first. One reasonable
choice is $\alpha_{\nu}^{0}=1/\sqrt{N}$

\begin{itemize}
\item In step (2) $\left(  N-1\right)  $ new basis states $\overline{a}%
_{i}^{\ast}$ $\left(  1\leq i\leq N-1\right)  $ are formed which are
normalized and orthogonal to each other and to $a_{0}^{\ast}$.

\item In step (3) the s-band Hamiltonian $H_{0}$ is constructed in this new
basis. One puts the state $a_{0}^{\ast}$ at the top so that its matrix
elements are $H_{0i}$ and $H_{i0}$.

\item In step (4) the $\left(  N-1\right)  $-sub Hamiltonian which does not
contain the state $a_{0}^{\ast}$ is diagonalized. This transforms the rest of
the basis $\left\{  \overline{a}_{i}^{\ast}\right\}  $ into a new basis
$\left\{  a_{0}^{\ast},a_{i}^{\ast}\right\}  $ (but keeps the state
$a_{0}^{\ast}$ unchanged). The resulting Hamilton matrix for the s-band then
has the form%
\begin{equation}
H_{0}=\left(
\begin{array}
[c]{ccccc}%
E(0) & V_{fr}(1) & V_{fr}(2) & ... & V_{fr}(N-1)\\
V_{fr}(1) & E(1) & 0 & ... & 0\\
V_{fr}(2) & 0 & E(2) & ... & 0\\
.. & ... & ... & ... & ...\\
V_{fr}(N-1) & 0 & 0 & ... & E(N-1)
\end{array}
\right)  \label{hmat}%
\end{equation}
The creation operators of the new basis are given by the set $\left\{
a_{0}^{\ast},a_{i}^{\ast}\right\}  ,$ ($0<i\leq N-1)$. The $a_{i}^{\ast}$ can
be expressed in terms of the s-states; $a_{i}^{\ast}=\sum_{\nu=1}^{N}%
\alpha_{\nu}^{i}c_{\nu}^{\ast}$. The state $a_{0}^{\ast}$ uniquely determines
the other states $a_{i}^{\ast}$. The state $a_{0}^{\ast}$ is coupled through
the matrix elements $V_{fr}\left(  i\right)  $ to the states $a_{i}^{\ast}$,
which makes the state $a_{0}^{\ast}$ an artificial Friedel resonance. The
matrix elements $E\left(  i\right)  $ and V$_{fr}\left(  i\right)  $ are given
as%
\begin{align*}
E(i)  &  =\sum_{\nu}\alpha_{\nu}^{i}\varepsilon_{\nu}\alpha_{\nu}^{i}\\
V_{fr}\left(  i\right)   &  =%
{\textstyle\sum_{\nu}}
\alpha_{\nu}^{0}\varepsilon_{\nu}\alpha_{\nu}^{i}%
\end{align*}

\item In the final step (5) the state $a_{0}^{\ast\text{ }}$is rotated in the
$N$-dimensional Hilbert space. In each cycle the state $a_{0}^{\ast}$ is
rotated in the $\left(  a_{0}^{\ast}\text{,}a_{i_{0}}^{\ast}\right)  $ plane
by an angle $\theta_{i_{0}}$ for $1\leq i_{0}\leq N-1$. Each rotation by
$\theta_{i_{0}}$ yields a new $\overline{a_{0}}^{\ast}$
\[
\overline{a_{0}}^{\ast}=a_{0}^{\ast}\cos\theta_{i_{0}}+a_{i_{0}}^{\ast}%
\sin\theta_{i_{0}}%
\]

The rotation leaves the whole basis $\left\{  a_{0}^{\ast},a_{i}^{\ast
}\right\}  $ orthonormal. Step (4), the diagonalization of the $\left(
N-1\right)  $-sub Hamiltonian, is now much quicker because the $\left(
N-1\right)  $-sub-hamiltonian is already diagonal with the exception of the
$i_{0}$- row and the $i_{0}$-column . For each rotation plane $\left(
a_{0}^{\ast}\text{,}a_{i_{0}}^{\ast}\right)  $ the optimal $a_{0}^{\ast}$ with
the lowest energy expectation value is determined. This cycle is repeated
until one reaches the absolute minimum of the energy expectation value. In the
example of the Friedel resonance Hamiltonian this energy agrees numerically
with an accuracy of $10^{-15}$ with the exact ground-state energy of the
Friedel Hamiltonian \cite{B91}. For the Kondo impurity the procedure is
stopped when the expectation value changes by less than $10^{-10}$ during a
full cycle.
\end{itemize}

\subsection{Matrix elements}

In equ.(\ref{Hsd}) we expressed the Kondo Hamiltonian in terms of field
operators. We can also express it in the basis $\left\{  c_{\nu}^{\ast
}\right\}  $. Here it has the form%
\[
H_{sd}=%
{\textstyle\sum_{\nu,\nu^{\prime}}}
J_{\nu,\nu^{\prime}}%
\begin{array}
[c]{c}%
\left(  S_{+}c_{\nu\downarrow}^{\dag}\left(  0\right)  c_{\nu^{\prime}%
\uparrow}\left(  0\right)  +S_{-}c_{\nu\uparrow}^{\dag}\left(  0\right)
c_{\nu^{\prime}\downarrow}\left(  0\right)  \right) \\
+S_{z}\left(  c_{\nu\uparrow}^{\dag}\left(  0\right)  c_{\nu^{\prime}\uparrow
}\left(  0\right)  -c_{\nu\downarrow}^{\dag}\left(  0\right)  c_{\nu^{\prime
}\downarrow}\left(  0\right)  \right)
\end{array}
\]
with
\[
J_{\nu,\nu^{\prime}}=J\sqrt{\frac{\Delta\zeta_{\nu}}{2}}\sqrt{\frac
{\Delta\zeta_{\nu^{\prime}}}{2}},\text{ }\Delta\zeta_{\nu}=\zeta_{\nu}%
-\zeta_{\nu-1}%
\]

We will, however, use (a slighly modified version of) equ.(\ref{Hsd}). The
non-spin-flip part can be rewritten as%
\begin{equation}
H_{nsf}=v_{a}JS_{z}\left(  -\Psi_{\uparrow}\left(  0\right)  \Psi_{\uparrow
}^{\dag}\left(  0\right)  +\Psi_{\downarrow}\left(  0\right)  \Psi
_{\downarrow}^{\dag}\left(  0\right)  \right)
\end{equation}
where we anti-commuted the field operators.

We consider first the example of the non-spin-flip part where the spin of the
d-impurity and the s-electrons are anti-parallel:
\begin{align*}
\left\langle \psi_{B}\left\vert H_{_{\uparrow\uparrow}}\right\vert \psi
_{B}\right\rangle  &  =-v_{a}J\left\langle a_{0\uparrow}^{\ast}d_{\downarrow
}^{\ast}\mathbf{0}_{a\uparrow}\mathbf{0}_{b\downarrow}\left\vert S_{z}%
\Psi_{\uparrow}\left(  0\right)  \Psi_{\uparrow}^{\dag}\left(  0\right)
\right\vert a_{0\uparrow}^{\ast}d_{\downarrow}^{\ast}\mathbf{0}_{a\uparrow
}\mathbf{0}_{b\downarrow}\right\rangle \\
&  =\frac{v_{a}J}{2}\left\langle a_{0\uparrow}^{\ast}d_{\downarrow}^{\ast
}\mathbf{0}_{a\uparrow}\mathbf{0}_{b\downarrow}\left\vert \Psi_{\uparrow
}\left(  0\right)  \Psi_{\uparrow}^{\dag}\left(  0\right)  \right\vert
a_{0\uparrow}^{\ast}d_{\downarrow}^{\ast}\mathbf{0}_{a\uparrow}\mathbf{0}%
_{b\downarrow}\right\rangle \\
&  =\frac{v_{a}J}{2}\left\langle \Psi_{\uparrow}\left(  0\right)
a_{0\uparrow}^{\ast}\mathbf{0}_{a\uparrow}d_{\downarrow}^{\ast}\mathbf{0}%
_{b\downarrow}\left\vert {}\right\vert \Psi_{\uparrow}^{\dag}\left(  0\right)
a_{0\uparrow}^{\ast}\mathbf{0}_{a\uparrow}d_{\downarrow}^{\ast}\mathbf{0}%
_{b\downarrow}\right\rangle \\
&  =\frac{v_{a}J}{2}\sum_{i=0}^{n}\left\vert A_{i}^{a}\right\vert ^{2}%
\end{align*}
where $A_{i}^{a}$ is the amplitude of the state $a_{i}^{\ast}$ at $r=0$. The
summation is from $0\leq i\leq n$. Similarly one obtains for the non-spin-flip
part where the spin of the d-impurity and the s-electrons are parallel:%
\[
\left\langle \psi_{B}\left\vert H_{_{\downarrow\downarrow}}\right\vert
\psi_{B}\right\rangle =v_{a}J\left\langle a_{0\uparrow}^{\ast}d_{\downarrow
}^{\ast}\mathbf{0}_{a\uparrow}\mathbf{0}_{b\downarrow}\left\vert JS_{z}%
\Psi_{\downarrow}\left(  0\right)  \Psi_{\downarrow}^{\dag}\left(  0\right)
\right\vert a_{0\uparrow}^{\ast}d_{\downarrow}^{\ast}\mathbf{0}_{a\uparrow
}\mathbf{0}_{b\downarrow}\right\rangle =-\frac{v_{a}J}{2}\sum_{i=1}%
^{n}\left\vert A_{i}^{b}\right\vert ^{2}%
\]
$A_{i}^{b}$ is the amplitude of the state $b_{i}^{\ast}$ at the origin. The
latter summation is from $1\leq i\leq n$.

As an example for the spin-flip matrices one obtains%

\begin{align*}
\left\langle \psi_{\overline{B}}\left\vert H_{_{\uparrow\downarrow}%
}\right\vert \psi_{B}\right\rangle =v_{a}J  &  \left\langle d_{\uparrow}%
^{\ast}a_{0\downarrow}^{\ast}\mathbf{0}_{b\uparrow}\mathbf{0}_{a\downarrow
}\left\vert S_{+}\Psi_{\downarrow}^{\dag}\left(  0\right)  \Psi_{\uparrow
}\left(  0\right)  \right\vert a_{0\uparrow}^{\ast}d_{\downarrow}^{\ast
}\mathbf{0}_{a\uparrow}\mathbf{0}_{b\downarrow}\right\rangle \\
&  =-v_{a}J\left\langle \Psi_{\uparrow}^{\dag}\left(  0\right)  d_{\uparrow
}^{\ast}a_{0\downarrow}^{\ast}\mathbf{0}_{b\uparrow}\mathbf{0}_{a\downarrow
}\left\Vert {}\right\Vert \Psi_{\downarrow}^{\dag}\left(  0\right)
a_{0\uparrow}^{\ast}d_{\uparrow}^{\ast}\mathbf{0}_{a\uparrow}\mathbf{0}%
_{b\downarrow}\right\rangle \\
&  =-v_{a}J\left\langle \Psi_{\uparrow}^{\dag}\left(  0\right)  \mathbf{0}%
_{b\uparrow}a_{0\downarrow}^{\ast}\mathbf{0}_{a\downarrow}\left\Vert
{}\right\Vert a_{0\uparrow}^{\ast}\mathbf{0}_{a\uparrow}\Psi_{\downarrow
}^{\dag}\left(  0\right)  \mathbf{0}_{b\downarrow}\right\rangle \\
&  =-v_{a}J\left\langle a_{0\downarrow}^{\ast}\mathbf{0}_{a\downarrow
}\left\vert {}\right\vert \Psi_{\downarrow}^{\dag}\left(  0\right)
\mathbf{0}_{b\downarrow}\right\rangle \left\langle \Psi_{\uparrow}^{\dag
}\left(  0\right)  \mathbf{0}_{b\uparrow}\left\vert {}\right\vert
a_{0\uparrow}^{\ast}\mathbf{0}_{a\uparrow}\right\rangle \\
&  =-v_{a}J\left\vert \left\langle \Psi^{\dag}\left(  0\right)  \mathbf{0}%
_{b}\left\vert {}\right\vert a_{0}^{\ast}\mathbf{0}_{a}\right\rangle
\right\vert ^{2}%
\end{align*}
The final brackets $\left\langle {}\right\rangle $ represent a multi-scalar
product which is given by the determinant of the following $\left(
n+1\right)  \times\left(  n+1\right)  $ matrix.
\begin{align*}
&  \left\langle \Psi^{\dag}\left(  0\right)  \mathbf{0}_{b}\left\vert
{}\right\vert _{0}^{\ast}\mathbf{0}_{a}\right\rangle \\
&  =\left\vert
\begin{array}
[c]{cccc}%
\left\langle \Psi^{\ast}\left(  0\right)  |a_{0}^{\ast}\right\rangle  &
\left\langle \Psi^{\ast}\left(  0\right)  |a_{1}^{\ast}\right\rangle  &  &
\left\langle \Psi^{\ast}\left(  0\right)  |a_{n}^{\ast}\right\rangle \\
\left\langle b_{1}^{\ast}|a_{0}^{\ast}\right\rangle  & \left\langle
b_{1}^{\ast}|a_{1}^{\ast}\right\rangle  &  & \left\langle b_{1}^{\ast}%
|a_{n}^{\ast}\right\rangle \\
&  &  & \\
\left\langle b_{n}^{\ast}|a_{0}^{\ast}\right\rangle  & \left\langle
b_{n}^{\ast}|a_{1}^{\ast}\right\rangle  &  & \left\langle b_{n}^{\ast}%
|a_{n}^{\ast}\right\rangle
\end{array}
\right\vert
\end{align*}%
\[
=\left\vert
\begin{array}
[c]{cccc}%
A_{0}^{a} & A_{1}^{a} & ... & A_{n}^{a}\\
\left\langle b_{1}^{\ast}|a_{0}^{\ast}\right\rangle  & \left\langle
b_{1}^{\ast}|a_{1}^{\ast}\right\rangle  & ... & \left\langle b_{1}^{\ast
}|a_{n}^{\ast}\right\rangle \\
... & ... & ... & ...\\
\left\langle b_{n}^{\ast}|a_{0}^{\ast}\right\rangle  & \left\langle
b_{n}^{\ast}|a_{1}^{\ast}\right\rangle  & ... & \left\langle b_{n}^{\ast
}|a_{n}^{\ast}\right\rangle
\end{array}
\right\vert
\]

\subsection{Energy optimization}

For a given set of bases $\left\{  a_{0}^{\ast},a_{i}^{\ast}\right\}  $ and
$\left\{  b_{0}^{\ast},b_{i}^{\ast}\right\}  $ the energy expectation value of
the Kondo state is given by
\[
\left\langle E\right\rangle =\frac{\left\langle \psi_{K}\left\vert
H\right\vert \psi_{K}\right\rangle }{\left\langle \psi_{K}|\psi_{K}%
\right\rangle }%
\]

The numerator of the expectation value of the ground-state energy is given by
\begin{equation}
\left\langle \psi_{K}\left\vert H\right\vert \psi_{K}\right\rangle =\left(
\begin{array}
[c]{cccc}%
B & \overline{C} & C & \overline{B}%
\end{array}
\right)  \left(
\begin{array}
[c]{cccc}%
H_{BB} & H_{B\overline{C}} & H_{BC} & H_{B\overline{B}}\\
H_{\overline{C}B} & H_{\overline{C}\overline{C}} & H_{\overline{C}C} &
H_{\overline{C}\overline{B}}\\
H_{CB} & H_{C\overline{C}} & H_{CC} & H_{C\overline{\overline{B}}}\\
H_{\overline{B}B} & H_{\overline{B}\overline{C}} & H_{\overline{B}C} &
H_{\overline{B}\overline{B}}%
\end{array}
\right)  \left(
\begin{array}
[c]{c}%
B\\
\overline{C}\\
C\\
\overline{B}%
\end{array}
\right)
\end{equation}
where $\left(  V\right)  =\left(
\begin{array}
[c]{cccc}%
B & \overline{C} & C & \overline{B}%
\end{array}
\right)  ^{\dag}.$ The denominator is given by%
\[
\left\langle \psi_{K}|\psi_{K}\right\rangle =\left(  V\right)  ^{\dag}\left(
N\right)  \left(  V\right)
\]
where $\left(  N\right)  $ is the non-diagonal symmetric matrix of the scalar
products $\left\langle \psi_{X}|\psi_{Y}\right\rangle $ (since the Slater
states $\psi_{X}$ are pair-wise not orthogonal)%
\[
\left(  N\right)  =\left(
\begin{array}
[c]{cccc}%
1 & \left\langle \psi_{B}|\psi_{\overline{C}}\right\rangle  & 0 & 0\\
\left\langle \psi_{\overline{C}}|\psi_{B}\right\rangle  & 1 & 0 & 0\\
0 & 0 & 1 & \left\langle \psi_{C}|\psi_{\overline{B}}\right\rangle \\
0 & 0 & \left\langle \psi_{\overline{B}}|\psi_{C}\right\rangle  & 1
\end{array}
\right)
\]
With an orthogonal transformation of $\left(  V\right)  $ the matrix $\left(
N\right)  $ can be transformed into a diagonal matrix $\left(  \overline
{N}\right)  =\left(  N_{ii}\delta_{ij}\right)  $. Replacing $V_{i}%
->V_{i}/\sqrt{\overline{N}_{ii}}$ transforms the matrix $\left(  N\right)  $
into the unity matrix.

Now we can require that the new vector $\left(  V^{\prime}\right)  $ is
normalized so that $\left\langle \psi_{K}|\psi_{K}\right\rangle =\left(
B^{\prime2}+\overline{C}^{\prime2}+C^{\prime2}+\overline{B}^{\prime2}\right)
=1$ and we have only to vary the numerator. This yields the eigenvalue
problem
\begin{equation}
\left(  H^{\prime}\right)  \left(  V^{\prime}\right)  =\Lambda\left(
V^{\prime}\right)  \label{ewp}%
\end{equation}
which can be easily solved numerically. The original coefficients $\left(
\begin{array}
[c]{cccc}%
B & \overline{C} & C & \overline{B}%
\end{array}
\right)  $ are obtained by reversing the transformations.


\newpage

\begin{thebibliography}{99}                                                                                               %


\bibitem {F28}J.Friedel, Philos.Mag. 43, 153 (1952); Adv.Phys. 3, 446 (1954);
Philos.Mag.Suppl. 7, 446 (1954); Can.J.Phys. 34, 1190 (1956); Nuovo Cimento
Suppl. 7, 287 (1958); J. Phys.Radium 19, 38 (1958)\newline

\bibitem {A31}P.W.Anderson, Phys.Rev. 124, 41 (1961)\newline

\bibitem {K8}J.Kondo, Prog.Theor.Phys. 32, 37 (1964) \newline

\bibitem {A51}P.W.Anderson, J.Phys. C3, 2436 (1970)\newline

\bibitem {W18}K.G.Wilson, Rev.Mod.Phys. 47, 773 (1975)\newline

\bibitem {F30}H.O.Frota and L.N.Oliveira, Phys.Rev. B33, 7871 (1986)\newline

\bibitem {N14}P.Nozieres, J.Low Temp.Phys. 17, 31 (1974)\newline

\bibitem {N5}P.Nozieres, Ann.Phys.Fr. 10, 19 (1985)\newline

\bibitem {N7}D.M.Newns and N.Read, Adv. in Phys. 36, 799 (1987)\newline

\bibitem {W12}P.B.Wiegmann, in Quantum Theory of Solids, edited by
I.M.Lifshits (MIR Publishers, Moscow, 1982), p. 238\newline

\bibitem {S29}P.Schlottmann, Phys.Reports 181, 1 (1989)\newline

\bibitem {H20}A.C.Hewson, The Kondo problem to heavy Fermions, Cambridge
University Press, 1993\newline

\bibitem {V7}C.M.Varma and Y.Yafet, Phys.Rev. B13, 2950 (1976)\newline

\bibitem {G19}O.Gunnarsson and K.Schoenhammer, Phys.Rev. B28, 4315
(1983)\newline

\bibitem {B103}N.E.Bickers, Rev.Mod.Phys. 59, 845 (1987)\newline

\bibitem {A50}N.Andrei, K.Furuya, and J.H.Lowenstein, Rev.Mod.Phys. 55, 331
(1983)\newline

\bibitem {B152}G.Bergmann, Phys.Rev. B 73, 092418 (2006) \newline

\bibitem {B151}G.Bergmann, Phys.Rev. B 74, 144420 (2006) \newline

\bibitem {G34}O.Gunnarsson and K.Schoenhammer, Phys.Rev. B31, 4815
(1985)\newline

\bibitem {B91}G.Bergmann, Z.Physik B102, 381 (1997)\newline

\bibitem {B92}G.Bergmann, Eur.Phys.J.B2, 233 (1998)\newline

\bibitem {S31}J.R.Schrieffer and P.A.Wolff, Phys.Rev. 149, 491 (1967)\newline

\bibitem {Y2}K.Yosida, Phys.Rev. 147, 223 (1966)\newline
\end{thebibliography}
\end{document}